\documentclass[twocolumn,showpacs,preprintnumbers,amsmath,amssymb]{revtex4}

\usepackage{graphicx}
\usepackage{dcolumn}
\usepackage{bm}
\usepackage[latin1]{inputenc}

\begin{document}

\title{Quantum breathers in capacitively coupled Josephson junctions:
Correlations, number conservation, and entanglement}
\author{R. A. Pinto and S. Flach}
\affiliation{Max-Planck-Institut f\"ur Physik komplexer Systeme, N\"othnitzer Str. 38, 01187 Dresden, Germany }

\date{\today}

\begin{abstract}
We consider the classical and quantum dynamics of excitations in a system of two
capacitively coupled Josephson junctions. In the classical case
the equations of motion admit discrete breather solutions, which are time periodic and
localized predominantly on one of the junctions. In the quantum case breather states are
found in the central part of the energy spectrum of the confined nonescaping states
of the system. We perform a systematic analysis of their tunneling frequency, 
site correlations, fluctuations of the number of quanta, and entanglement.
Quantum breather states 
show strong site correlation of quanta and are characterized by a
strong excitation of quanta on one junction
which perform slow coherent tunneling motion from one junction to the
other. They suppress fluctuations of
the total number of excited quanta.
Quantum breather states are the least entangled states among the group
of eigenstates in the same range of the energy spectrum. 
We describe how quantum breather excitations could be experimentally observed by employing 
the already developed techniques
for quantum information processing using Josephson junctions. 

\end{abstract}


\maketitle

\section{Introduction}

Josephson junctions are the subject of extensive studies in
quantum information experiments because they possess two attractive properties:
In their classical regime they are nonlinear devices,
but also show macroscopic quantum behavior \cite{Likharev,Leggett,Esteve}.
The dynamics of a biased
Josephson junction (JJ) is analogous to the dynamics of a particle with a mass
proportional to the junction capacitance $C_J$,
moving on a tilted washboard potential
\begin{equation}
U(\varphi) = -I_c\frac{\Phi_0}{2\pi}\cos\varphi - I_b\varphi\frac{\Phi_0}{2\pi},
\end{equation}
which is sketched in Fig.\ref{fig1}-b.
Here $\varphi$ is the phase difference between the macroscopic wave functions
in both superconducting electrodes of the
junction, $I_b$ is the bias current, $I_c$ is the critical current of the junction, and $\Phi_0=h/2e$ the
flux quantum. When the energy of the particle is large enough to overcome the barrier
$\Delta U$ (that depends on the bias current $I_b$) it escapes and moves down the
potential, switching the junction into a resistive state with a nonzero voltage
proportional to $\dot{\varphi}$ across it.
Quantization of the system leads to discrete energy levels inside the
potential wells, which are nonequidistant because of the anharmonicity. Note that even if there is not
enough energy to classically overcome the barrier, the
particle may perform a quantum escape and tunnel outside the well, thus switching the junction into the resistive
state \cite{Likharev}. Thus each state inside the well is characterized by a bias and state-dependent
lifetime, or its inverse ---the escape rate.

Progress on manipulation of quantum JJs includes
spectroscopic analysis, better isolation schemes, and simultaneous measurement techniques
\cite{Leggett,Esteve,Steffen2006,Martinis,Steffen,Martinis2}, and
paves the way for using them as Josephson-junction qubits
in arrays for experiments on processing quantum information.
Typically the first two or three quantum levels of one
junction are used as quantum bits. Since the levels are nonequidistant, they
can be separately excited by applying microwave pulses.

However, improvements in experiments
manipulating Josephson-junction qubits may have applications beyond the
processing of quantum information. Operating the junctions at larger energies
in the quantum regime may
give rise to other interesting phenomena that nowadays can be experimentally observed by using already
developed experimental techniques. For instance, it was suggested that JJs operating at high energies may be used for experiments on quantum chaos
\cite{Graham,Montangero,Pozzo}.

Another interesting phenomenon is the excitation of discrete breathers. They are time periodic space
localized excitations in anharmonic lattices with translational
invariance \cite{physicstoday,FlachPhysRep295,Sievers,AubryPhysicaD103}. They localize energy exponentially
in space for short-range coupling between lattice sites, and have been experimentally observed in such
different systems as bond excitations in molecules, 
lattice vibrations and spin excitations in solids, electronic currents in coupled
JJs, light propagation in interacting optical waveguides, 
cantilever vibrations in micromechanical
arrays, cold atom dynamics in Bose-Einstein condensates loaded on optical lattices,
among others \cite{SchwarzPRL83,SatoNature,SwansonPRL82,TriasPRL84,BinderPRL84,EisenbergPRL81,
FleischerNature422,SatoPRL90,EiermannPRL92}.

In the quantum regime, quantum breathers (QB) \cite{Fleurov,ScottPhysLettA119,BernsteinNonlin3,
BernsteinPhysicaD68,WrightPhysicaD69,Eilbeck94,Wang,Aubry,Flach1,AubryPhysicaD103,Fleurov1998,kalosakas2,
Dorignac2004,Eilbeck2004,Pinto,Schulman,Schulman2006,Proville2006,Ivic2006} appear as
nearly degenerate many-quanta bound states. Though being extended
in a translationally invariant system, they are characterized by exponentially localized correlation functions
in full analogy to their classical counterparts \cite{Wang,Eilbeck03}. When such states
superpose the result is a spatially localized excitation with a very long time to tunnel from one lattice site to
another. 
At variance to the classical case,
the evolution of these excitations in time has not been experimentally studied in detail. So
far they have been indirectly observed by spectroscopic
analysis in molecules and solids \cite{Fillaux1990,Fillaux1998,Richter1988,GuyotSionnest1991,Dai1994,Chin1995,Jakob1996,
JakobPr75,Okuyama2001,Edler2004}.

The possibility to directly observe QB excitations evolving in time was
addressed by us in a letter \cite{Pinto2007} for a system of two capacitively
coupled JJs, where by calculating the eigenstates and the spectrum of the
system we identified QB
states as weakly splitted tunneling pairs of states
\cite{Aubry,Flach1,Fleurov1998,Pinto}. 
These eigenstates appear in the middle of the energy spectrum
of the system and are characterized by correlations between the two junctions -
if one of them is strongly excited, the other one is not, and vice versa.
By exciting one of the junctions to a large energy (many quanta), we strongly overlap with
QB tunneling states. Consequently we may trap the excitation on the initially
excited junction on a time scale which sensitively depends on the amount of energy excited,
and on the applied bias. We described how QB excitations could be directly
observed in time using the available techniques for manipulating JJs in
the quantum regime.

In this work we present an extended analysis of the system, 
performing a systematic and comparative analysis of different properties of
QB states. 
We study their tunneling rates, the site correlations of excited quanta,
the fluctuation of the number of excited quanta, and the entanglement
of the QB states. 

In section II we describe the model for the two coupled JJs
\cite{Berkley,Johnson,Blais} and
briefly consider the classical dynamics, where the equations of motion are
numerically solved finding discrete breather solutions. In section III we
consider the quantum model and introduce the basis we use to numerically diagonalize the
Hamiltonian matrix. We define correlations
functions which, together with the energy spectrum, will help us to identify
QB states. Then we compute the time evolution of initially localized excitations
and relate it to the spectral properties of the system. In section IV we address
the fluctuation of the total number of quanta 
in the eigenstates. In section V we
explore the entanglement of the eigenstates. In section VI we describe how
QB excitations evolving in time could be experimentally
observed, and discuss how escaping and decoherence (effects that are not taken into
account in the quantum model) would affect the observations. We conclude in section VII.

\section{The model and classical dynamics}

\begin{figure}[h]
\includegraphics[width=1.15in]{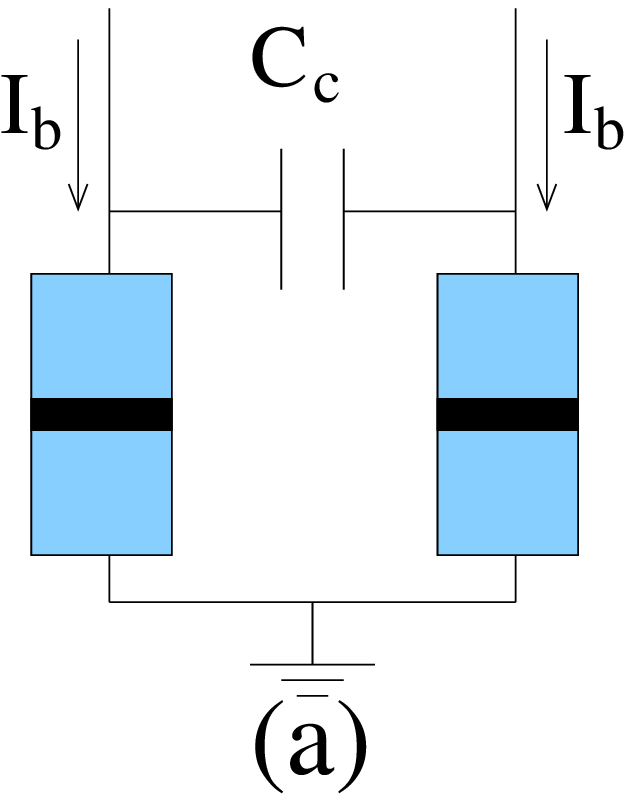}
\hspace{0.1cm}
\includegraphics[width=1.0in]{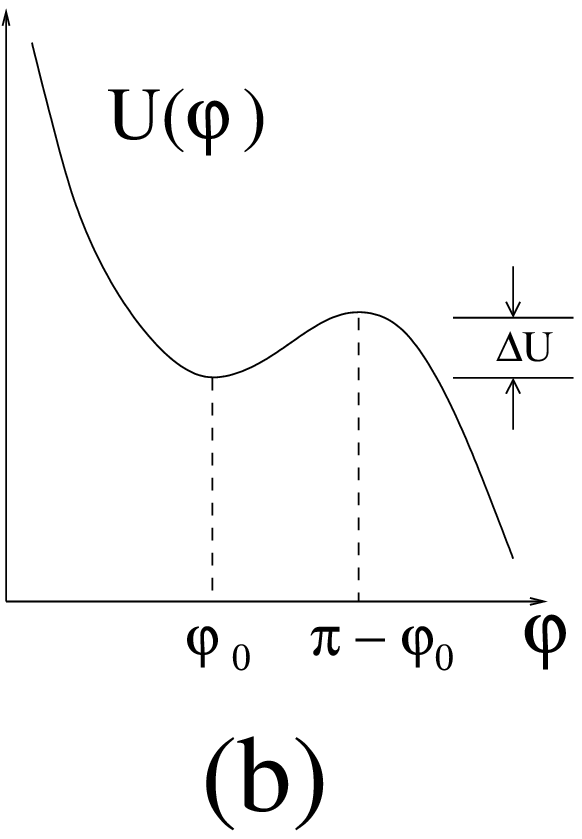}
\hspace{0.2cm}
\includegraphics[width=1.9in]{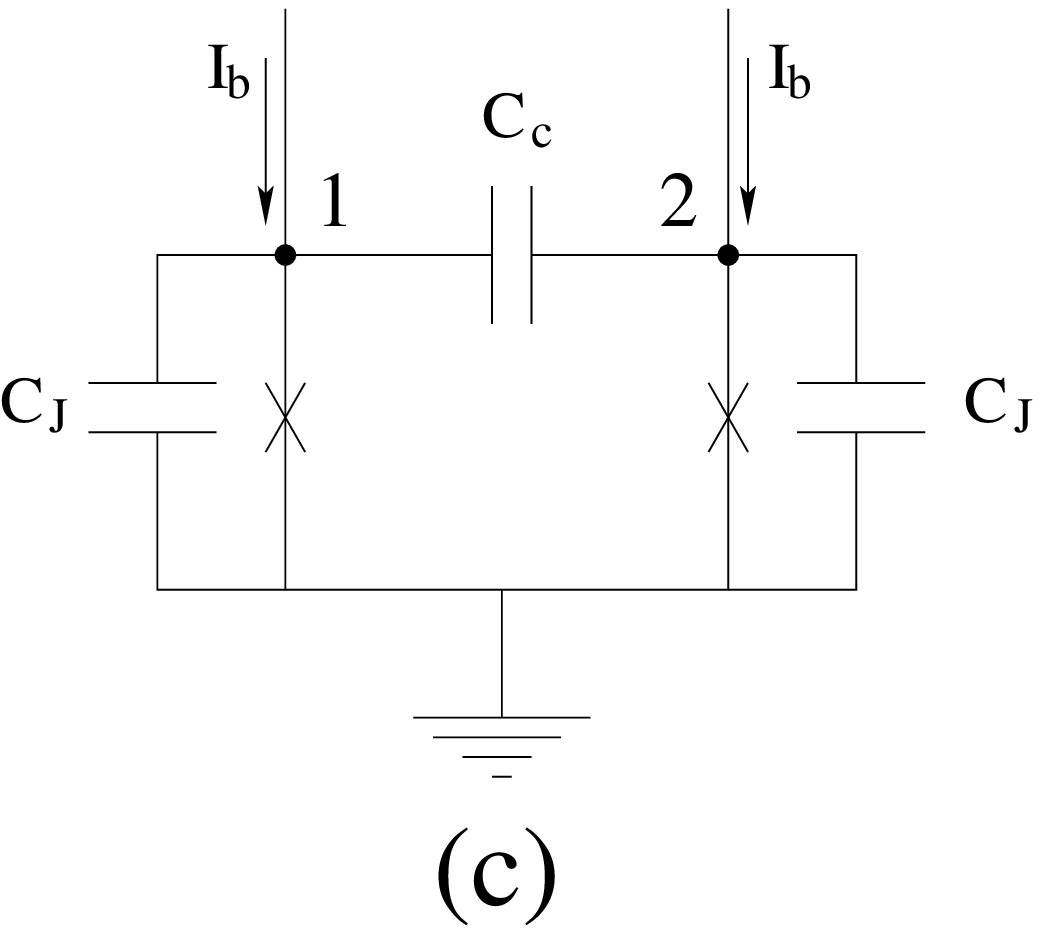}
\caption{\label{fig1}(a) Sketch of the two capacitively coupled Josephson
junctions. (b) Sketch of the washboard potential for a single
current-biased Josephson-junction. (c) Circuit diagram for two ideal capacitively coupled
Josephson junctions.}
\end{figure}

\begin{figure}[h]
\includegraphics[width=2.0in]{Figure2_a.eps}
\includegraphics[width=1.2in]{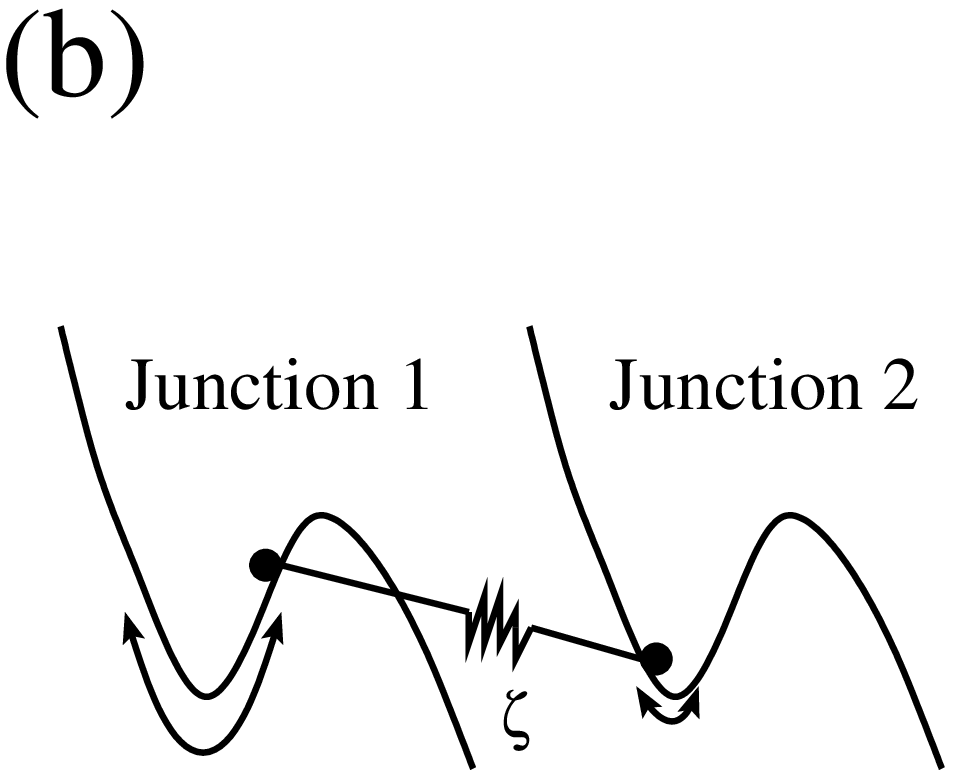}
\caption{\label{db}(a) Time evolution of the phase differences, and (b) corresponding sketch, of an
exact discrete breather solution of the equations
of motion (\ref{eq:emotion}) with frequency $\Omega_b=0.303\omega_p(0)$ (time is measured in units of the inverse
plasma frequency at zero bias $\omega_p(0)^{-1}$). The parameters are
$\gamma=0.99$ and $\zeta=0.1$.}
\end{figure}

The system is sketched in Fig.\ref{fig1}-a: two JJs are coupled by a
capacitance $C_c$, and they are biased by the same current $I_b$. The strength
of the coupling due to the capacitor is $\zeta =
C_c/(C_c+C_J)$.
The Hamiltonian of the system is
\begin{equation}
H = \frac{P_1^2}{2m}+ \frac{P_2^2}{2m} + U(\varphi_1)+U(\varphi_2) + \frac{\zeta}{m}P_1P_2,
\end{equation}
where
\begin{eqnarray}
m &=& C_J(1+\zeta)\left(\frac{\Phi_0}{2\pi}\right)^2 ,\\
P_{1,2} &=& (C_c+C_J)\left(\frac{\Phi_0}{2\pi}\right)^2 (\dot{\varphi}_{1,2} -
\zeta\dot{\varphi}_{2,1}).
\end{eqnarray}
Note that the conjugate momenta $P_{1,2}$ are proportional to the charge at
the nodes of the circuit (which are labeled in Fig.\ref{fig1}-c).
When the junctions are in the superconducting state, they behave like
two coupled anharmonic oscillators with plasma frequency $\omega_p(\gamma) =
\sqrt{2\pi I_c/\Phi_0 C_J(1+\zeta)} [1-\gamma^2]^{1/4}$, $\gamma =
I_b/I_c$ being the normalized bias current.
The classical equations of motion are given by
\begin{equation}\label{eq:emotion}
\ddot{\varphi}_{1,2} = -\frac{\Phi_0}{2\pi m}(\sin\varphi_{1,2} +
\zeta\sin\varphi_{2,1}) + \frac{\Phi_0}{2\pi m}(1+\zeta)\gamma\;.
\end{equation} 
Despite being invariant under permutation of the junction labels, these equations admit
discrete breather solutions \cite{FlachPhysRep295}, which are time periodic and for which the energy is
localized predominantly on one
of the junctions (Fig.\ref{db}). These orbits can be numerically computed with high accuracy using Newton
algorithms \cite{Flach2,Aubry1}.

The existence of discrete breathers is possible because the anharmonicity in the JJ potentials makes the frequency of these excitations (and all of their harmonics) nonresonant with the normal modes $\omega_{\pm} = \sqrt{1\pm\zeta}\omega_p(\gamma)$ of the coupled-junctions system, whose corresponding orbits are delocalized \cite{FlachPhysRep295}. For the parameters $\gamma=0.99$ and $\zeta=0.1$, the normal-mode frequencies are $\omega_+=0.394\omega_p(0)$ (in-phase mode) and $\omega_-=0.356\omega_p(0)$ (out-of-phase mode). The periodic solution shown in Fig.\ref{db} has a frequency below the out-of-phase mode frequency, thus the discrete breather solution is out of phase as well.

\section{Quantum dynamics: Exciting quantum breather states}\label{qdynam}

In the quantum case we compute the energy eigenvalues and the eigenstates of the system.
Since we are interested only in the energy transfer between the junctions, we
neglect quantum escape for states which will not escape in the 
classical limit. Thus we use a changed potential energy for the single
JJ by adding a hard wall
which prevents escape:
\begin{equation}
U_{q}(\varphi) = \left \{ \begin{array}{ll}
U(\varphi) & \textrm{if $\varphi \leq \pi -\varphi_0$} \\
\infty & \textrm{if $\varphi > \pi -\varphi_0$}
\end{array} \right.,
\end{equation}
where $\varphi_0 = \arcsin\gamma$ is the position of the minimum of the
potential and $\pi -\varphi_0$ gives the position of the first maximum to the right from
the equilibrium position $\varphi_0$ (Fig.\ref{fig1}-b).
We will later compare the obtained tunneling times with the true state dependent escape times.

The Hamiltonian of the two-junctions system is given by
\begin{equation}\label{eq:qHamiltonian}
\hat{H} = \hat{H}_1 + \hat{H}_2 + \zeta \hat{V},
\end{equation}
where $\hat{H}_i = \hat{P}_i^2/2m + U_q(\hat{\varphi}_i)$ is the single-junction Hamiltonian
and $\hat{V}=\hat{P}_1\hat{P}_2/m$ is the interaction that couples the junctions.
The eigenvalues $\varepsilon_{n_i}$ and eigenstates $|n_i\rangle$ of the
single-junction Hamiltonian $\hat{H}_i$ were
computed by using the Fourier grid Hamiltonian method \cite{Fourier}. 
$|n_i\rangle$ is also an eigenstate of the number operator
$\hat{n}_i$ with eigenvalue $n_i$. In the
harmonic approximation \cite{harmonic}
\begin{equation}\label{eq:number}
\hat{n}_i=\hat{a}_i^{\dagger}\hat{a}_i,
\end{equation}
where $\hat{a}_i^{\dagger}$ and $\hat{a}_i$ are
the bosonic creation and annihilation operators.
Since only states with energies below the classical escape energy (barrier) are taken into account,
the computed spectra have a finite upper bound.
The perturbation $\hat{V}$ does not conserve the total number of
quanta $n_1+n_2$, as seen from the dependence of the 
momentum operators 
on the bosonic creation and annihilation operators in the harmonic
approximation:
\begin{eqnarray}\label{eq:momenta}
\hat{P}_{1,2}= (\Phi_0/2\pi)\sqrt{(1+\zeta)C_J\hbar\omega_p/2} \nonumber \\
\times (\hat{a}_{1,2}-\hat{a}_{1,2}^{\dagger})/i.
\end{eqnarray}

The Hamiltonian matrix is written in the basis of product states of the
single-junction problem $\{|n_1,n_2\rangle = |n_1\rangle \otimes
|n_2\rangle\}$. The invariance of the Hamiltonian 
under permutation of the junction labels allows us to use symmetric and
antisymmetric basis states
\begin{equation}\label{eq:basis}
|n_1,n_2\rangle_{S,A} = \frac{1}{\sqrt{2}}(|n_1,n_2\rangle \pm |n_2,n_1\rangle )
\end{equation}
to reduce the full Hamiltonian matrix to two smaller symmetric and antisymmetric 
decompositions of $\hat{H}$, which after diagonalization respectively give the symmetric and
antisymmetric eigenstates of the system.

In order to identify quantum breather states, whose
corresponding classical orbits are characterized by energy localization,
we define the correlation functions:
\begin{equation}
f_{\mu}(1,2) = \langle\hat{n}_1\hat{n}_2\rangle_{\mu}
\end{equation}
\begin{equation}
f_{\mu}(1,1) = \langle\hat{n}_1^2\rangle_{\mu},
\end{equation}
where $\langle\hat{A}\rangle_{\mu} =
\langle\chi_{\mu}|\hat{A}|\chi_{\mu}\rangle$,
$\{|\chi_{\mu}\rangle\}$ being the set of eigenstates of the system.
The ratio $0 \leq f_{\mu}(1,2)/f_{\mu}(1,1) \leq 1$ measures the site correlation of quanta:
it is small when
quanta are site-correlated (i.e. when 
many quanta are located on one junction there are almost none on the other one) 
and close to one otherwise.

\begin{figure}[!t]
\includegraphics[width=3.3in]{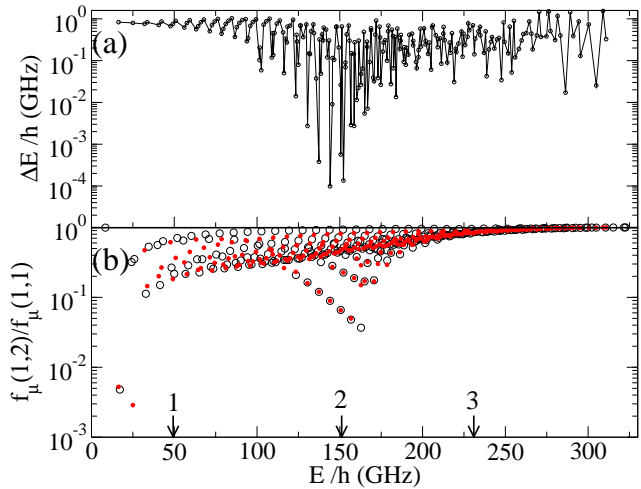}
\caption{\label{fig2}(a) Energy splitting and (b) correlation function
vs. energy of the eigenstates of the two-junctions system (open circles, symmetric
eigenstates; filled circles, antisymmetric eigenstates). The labeled arrows mark the energy
corresponding to the peak of the spectral intensity in Fig.\ref{fig3}-b, d, and f (see text).
The parameters are
$\gamma=0.945$ and $\zeta=0.1$ (22 levels per junction).}
\end{figure}

In Fig.\ref{fig2} we show the nearest neighbor energy spacing 
(tunneling splitting) and the correlation function of
the eigenstates. For this, and all the rest, we used
$I_c=13.3 \;\mu$A, $C_J=4.3$ pF, and $\zeta=0.1$, which are typical values in experiments.
We see that in the central part of the spectrum the energy splitting becomes
small in comparison to the average. The corresponding pairs of eigenstates, which are tunneling pairs,
are site correlated, and thus QBs. In these states
many quanta are localized on one junction and the tunneling time of
such an excitation from one junction to the other (given by the inverse energy
splitting between the eigenstates of the pair) can be exponentially large and depend
sensitively on the number of quanta excited.

Note that the tunneling of quanta between the JJs occurs without an obvious potential energy barrier being present (the interaction between the junctions is only through their momenta, as seen in the Hamiltonian (\ref{eq:qHamiltonian})). This process has been coined {\it dynamical tunneling} \cite{Davis1981JChemPhys75,Keshavamurthy2007IntRevPhysChem26,Keshavamurthy2005JChemPhys122}, to distinguish from the usual tunneling through a potential barrier. In dynamical tunneling, the barrier ---a so-called invariant separatrix manifold--- is only visible in phase space, where it separates two regions of regular classical motion between which the tunneling process takes place. Therefore, when referring to the tunneling between the JJs, we implicitly mean that it is dynamical.

The fact that the strongest site correlated eigenstates occur in the
central part of the energy spectrum may be easily explained as follows: Let $N$ be
the highest excited state in a single junction, with a corresponding
maximum energy $\Delta U$ (Fig.\ref{fig1}). For
two junctions the energy of the system with both junctions in the
$N$-th state is $2\Delta U$, which roughly is the width of the full
spectrum. Thus states of the form $|N,0\rangle$ and $|0,N\rangle$ that
have energy $\Delta U$ are located approximately in the middle.

\begin{figure}[!t]
\includegraphics[width=3.2in]{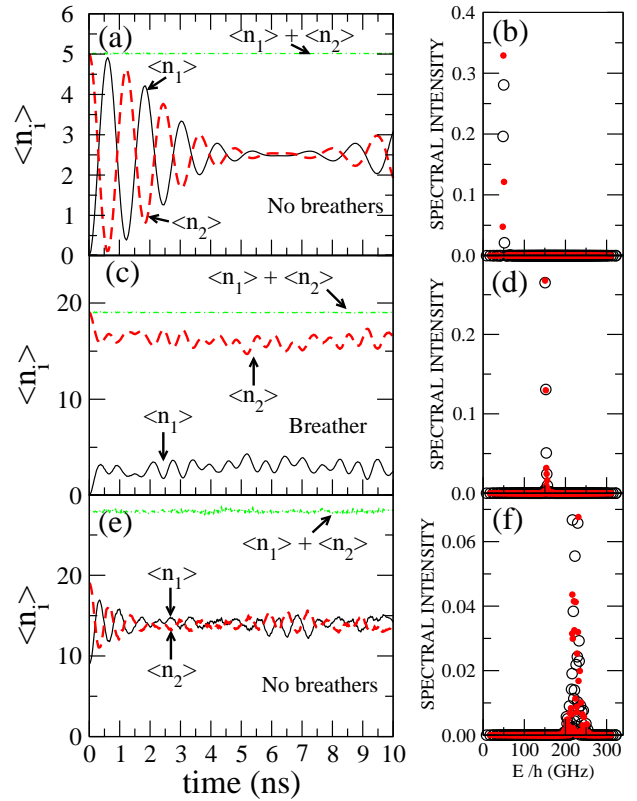}
\caption{\label{fig3}Time evolution of expectation values of the number of
quanta at each junction (left panels) for different initial excitations with corresponding spectral intensities (right panels). (a) and (b): $|\Psi_0\rangle =|0,5\rangle$;
(c) and (d): $|\Psi_0\rangle =|0,19\rangle$;
(e) and (f): $|\Psi_0\rangle =|9,19\rangle$. Open circles, symmetric
eigenstates; filled circles, antisymmetric eigenstates. The energies of the peaks in the spectral
intensity are marked
by labeled arrows in Fig.\ref{fig2}-b (see text). The parameters are
$\gamma=0.945$ and $\zeta=0.1$ (22 levels per junction).}
\end{figure} 

Having the eigenvalues and eigenstates, we compute the time evolution of
different initially localized excitations, and the expectation value of
the number of quanta at each junction $\langle \hat{n}_i\rangle (t)=
\langle\Psi(t)|\hat{n}_i|\Psi(t)\rangle$. Results are shown in
Fig.\ref{fig3}a, c, and e. We also compute the spectral intensity
$I_{\mu}^0 = |\langle \chi_{\mu}|\Psi_0\rangle|^2$, which
measures the strength of overlap of the initial state $|\Psi_0\rangle$ with the
eigenstates.
Results are shown in Fig.\ref{fig3}-b,
d, and f, where we can see a peak in each case, which corresponds to the 
arrows in Fig.\ref{fig2}-b.
We
can see that the initial state $|\Psi_0\rangle=|0,5\rangle$ overlaps with
eigenstates with an energy
splitting between them being relatively large and hence the
tunneling time of the initially localized excitation is short.
For the case $|\Psi_0\rangle=|0,19\rangle$ QBs are excited: The
excitation overlaps strongly with tunneling pairs of eigenstates in the
central part of the spectrum, which are
site correlated and nearly degenerate. The
tunneling time of such an excitation is very long, and thus keeps the quanta localized 
on their initial excitation site for corresponding
times.
Finally the initial state $|\Psi_0\rangle = |9,19\rangle$ overlaps with weakly site correlated
eigenstates with large energy
splitting. Hence the tunneling time is short again.

We computed also the time evolution of the expectation values of the number of
quanta for initial conditions which are coherent or incoherent (mixtures) superpositions
of product basis states with equal weights. This is relevant for experiments,
since it may be hard to excite one junction
to a determined state but easier to excite several states of the junction at the
same time. 
We used coherent superpositions (characterized by well defined 	states
$|\Psi_0\rangle$), and mixtures (characterized by their corresponding density
operators $\hat{\rho}_0$), of four basis states around the
already used initial states: For the 
state $|0,5\rangle$ we superposed the basis
states $|0,5\rangle$, $|0,6\rangle$, $|0,7\rangle$, and $|0,8\rangle$,
for $|0,19\rangle$ the
basis states $|0,20\rangle$, $|0,19\rangle$, $|0,18\rangle$, and $|0,17\rangle$, and
for $|9,19\rangle$ the
basis states $|9,20\rangle$, $|9,19\rangle$, $|9,18\rangle$, and $|9,17\rangle$.
Both for superposition and mixture of basis states, the results
are qualitatively similar to those shown in Fig.\ref{fig3}. Therefore we
expect that some imprecision in exciting an initial state in the junctions
would not affect in a relevant way the results. we may also conclude, that
the excitation of QB states does not rely on the phase coherence.
That conclusion will be supported later by the study of entanglement.

\begin{figure}[!t]
\begin{center}
\includegraphics[width=2.7in]{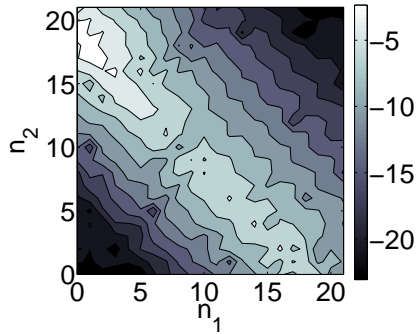}
\caption{\label{dist1}Contour plot of the logarithm of the density of the asymmetric state $|\chi\rangle =
(|\chi_{b}^{(S)}\rangle + |\chi_{b}^{(A)}\rangle)/\sqrt{2}$ as a function
of the number of quanta at junctions 1 and 2 (see text). The parameters are
$\gamma=0.945$ and $\zeta=0.1$ (22 levels per junction).}
\end{center}
\end{figure}

Let us estimate how many quanta should be excited in the junctions in order to obtain
QBs (tunneling pairs).  We compute the density
$\rho(n_1,n_2)=|\langle n_1,n_2|\chi\rangle|^2$ of the asymmetric state  $|\chi\rangle =
(|\chi_{b}^{(S)}\rangle + |\chi_{b}^{(A)}\rangle)/\sqrt{2}$, where 
$|\chi_{b}^{(S,A)}\rangle$ are the eigenstates belonging to a tunneling
pair \cite{Pinto}.
In Fig. \ref{dist1} we show a contour plot of the logarithm of the density for the 
tunneling pair with energy
marked by the arrow labeled by number two in Fig. \ref{fig2}-(b). We see
that the density has its maximum around $n_1=19$ and $n_2=0$ which is consistent
with the result shown in Fig. \ref{fig3}-c and d where QBs were excited
by using this combination of quanta in the junctions.

\section{Fluctuations of the total number of quanta}

\begin{figure}[!t]
\includegraphics[width=3.4in]{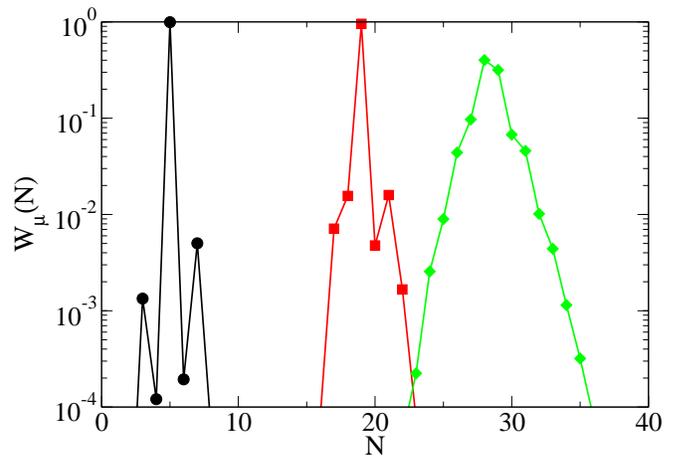}
\caption{\label{weight}From Left to right: Weight function as a function of the total number of
quanta $N=n_1+n_2$ for the three symmetric eigenstates at the peaks of the spectral intensities shown in Fig.\ref{fig3}-b, d, and f respectively. The parameters are
$\gamma=0.945$ and $\zeta=0.1$ (22 levels per junction).}
\end{figure}

Even though the Hamiltonian does not commute with the
total number of quanta $\hat{N}=\hat{n}_1 + \hat{n}_2$, in Fig.\ref{fig3}-a,
c and e we see that
its expectation value has very small fluctuations (less than
one). We can see this approximate conservation of the number of quanta 
also in the density plotted in
Fig.\ref{dist1}, where shows a rim along the
line $n_1+n_2 = N$ (=19). This might be computationally advantageous when considering larger
systems because the strict conservation of $N=n_1+n_2$ would allow us to
truncate the Hilbert space and work within a subspace with fixed $N$.
Each time the interaction operator $\hat{V}$ acts on a basis state with given $N$,
it will generate also states with $N\pm 2$, as can be seen from first order
perturbation theory in $\zeta$ at the harmonic
approximation. 
To study these fluctuations numerically we computed the following
weight function for each eigenstate, which measures the relative contribution
of all basis states with a given $N$ to the eigenstate under consideration:
\begin{equation}
W_{\mu}(N) = \sum_{\substack{n_1,n_2 \\ n_1+n_2=N}}|\langle n_1,n_2|\chi_{\mu}\rangle|^2
\end{equation}
In Fig.\ref{weight} we show the weight function for the three symmetric
eigenstates which correspond to the peaks of the spectral intensities shown in
Fig. \ref{fig3}-b, d, and f respectively.
For the lowest-energy state we can see the expected appearance of two 
satellite peaks separated by two
quanta from the central one. For the higher-energy eigenstates 
the harmonic approximation does not hold. Most importantly we see, that
for states in the lower and middle part of the energy spectrum, the fluctuation
of the number of quanta is weak, and corresponding states contribute less than one percent
to the eigenstate. This is apparently not true at the upper end of the energy spectrum.

Note that the obtained amplitude of fluctuations in the time evolution is much less due
to averaging effects and the smallness of the strength of the perturbation $\zeta$. The calculation
of $\langle \hat{N}\rangle(t)$ from perturbation theory in the harmonic
approximation shows that this quantity oscillates in time with an amplitude
that is proportional to $\zeta^2 n_0m_0$, where $n_0$, and
$m_0$ are the energy levels initially excited in the junctions. For $\zeta=0.1$
one finds that the fluctuations are of the order of $10^{-2}$ in the case
shown in Fig.\ref{fig3}-a; and of the order of $10^{-1}$ in the cases in
Fig.\ref{fig3}-c and e. Numerical results are consistent with these estimates.   

\begin{figure}[!t]
\includegraphics[width=3.4in]{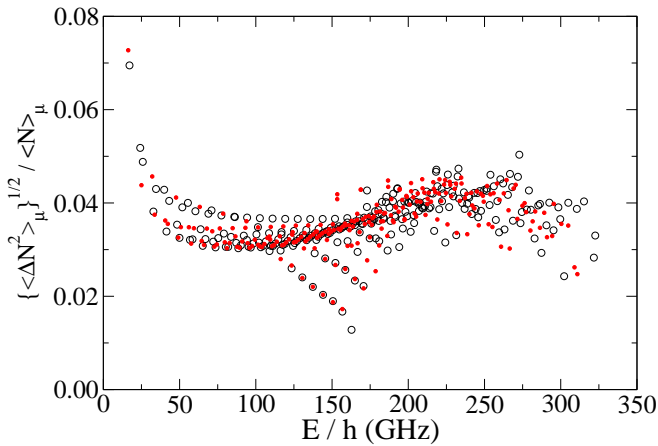}
\caption{\label{fluctN}Fluctuation of the total number of quanta for the
eigenstates of the coupled-junction system (open circles, symmetric
eigenstates; filled circles, antisymmetric eigenstates). The parameters are
$\gamma=0.945$ and $\zeta=0.1$ (22 levels per junction).}
\end{figure}

To characterize the variation in the total number of
quanta in the eigenstates we computed the fluctuation:
\begin{equation}\label{eq:fluctN}
\sqrt{\langle \Delta N^2\rangle_{\mu}} = \sqrt{\langle \hat{N}^2\rangle_{\mu} - \langle
\hat{N}\rangle^2_{\mu}}
\end{equation}
In Fig.\ref{fluctN} we plot the relative fluctuation $\sqrt{\langle\Delta
N^2\rangle_{\mu}}/\langle\hat{N}\rangle_{\mu}$ for the eigenstates, where we
can see that it is very small, and for the QB states in the central part of the
spectrum it has the smallest values.
This follows from the fact that QBs are close to eigenstates having the form
\begin{equation}\label{eq:QB}
|\chi\rangle_{QB} \simeq \frac{1}{\sqrt{2}}(|n,0\rangle \pm |0,n\rangle ),
\end{equation}
with $n\lesssim N$ ($N$ is the number of levels per junctions). These are eigenstates of the total number operator $\hat{N}=\hat{n}_1+\hat{n}_2$, for which the corresponding fluctuations given by (\ref{eq:fluctN}) vanish.

\section{Entanglement of QB states}

Since QB states are close to eigenstates of the $\zeta=0$ case given by (\ref{eq:QB}),
one expects that the degree of entanglement in QB
states is similar to the degree of entanglement in such states.
Since only two basis states are involved, it can not be a state of maximum
entanglement.   
We measured the degree of entanglement in the eigenstates of the system by
minimizing the distance of a given state to the 
space of product states, which depends on the largest eigenvalue
of the reduced density matrix
\cite{geoent1,geoent2,geoent3}:
\begin{equation}
\Delta = \sum_{n_1,n_2}^N(\chi_{n_1,n_2}^{\mu} - f_{n_1}g_{n_2})^2,
\end{equation}
where $\chi_{n_1,n_2}^{\mu}=\langle n_1,n_2|\chi_{\mu}\rangle$, and the
functions $f_{n_1}$ and $g_{n_2}$ are such that $\Delta$ is minimum (see
appendix for explicit formulas).
$\Delta$ measures how far a given eigenstate of
the two-junction system is from being a product of single-junction
states, and has values $0<\Delta<1$ (see appendix).
For $\zeta=0$ the
eigenstates of the system are the basis states given by eq. (\ref{eq:basis}), where for
$n_1=n_2$ it follows that $\Delta=0$, and for $n_1\neq n_2$ (which includes the state
in eq. (\ref{eq:QB})) $\Delta = 0.5$.
This measure has a direct relation to the distance of a given eigenstate
from a possible one obtained after performing a Hartree approximation \cite{geoent1}.

In a quantum integrable model with two degrees of freedom \cite{Aubry} it was shown
that the region of existence of QB states in the energy spectrum is separated
from other states by the energy threshold for which
discrete breathers exist in the corresponding classical model. Pairs of eigenstates with energies
beyond this threshold show exponentially decreasing energy splitting. In
a similar quantum model \cite{Tonel2005,Fu2006}, it was shown that at the mentioned energy
threshold the entanglement (using the von Neumann entropy) becomes maximum and then decreases with
energy. From these two results we expect that QB states in our case 
show decreasing entanglement $\Delta$ with
energy, tending to 0.5.
\begin{figure}[!t]
\includegraphics[width=3.4in]{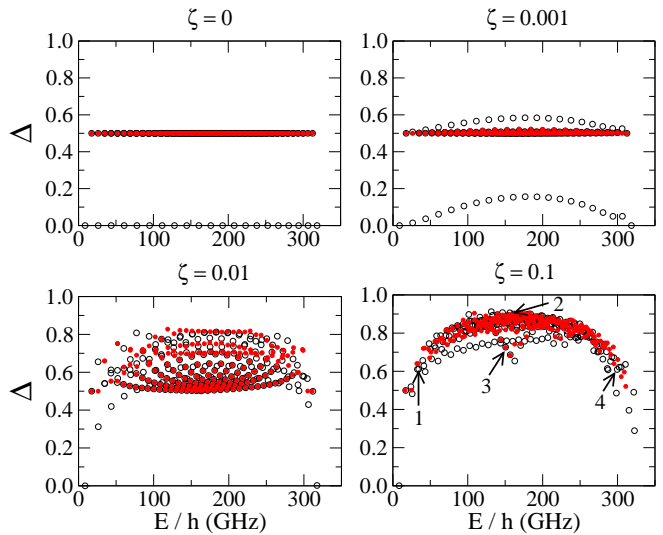}
\caption{\label{entanglement}Entanglement of the
eigenstates of the coupled-junction system for different values of the
coupling strength $\zeta$ (open circles, symmetric
eigenstates; filled circles, antisymmetric eigenstates). Here $\gamma=0.945$.}
\end{figure} 
\begin{figure}[!ht]
\begin{center}
\includegraphics[width=1.68in]{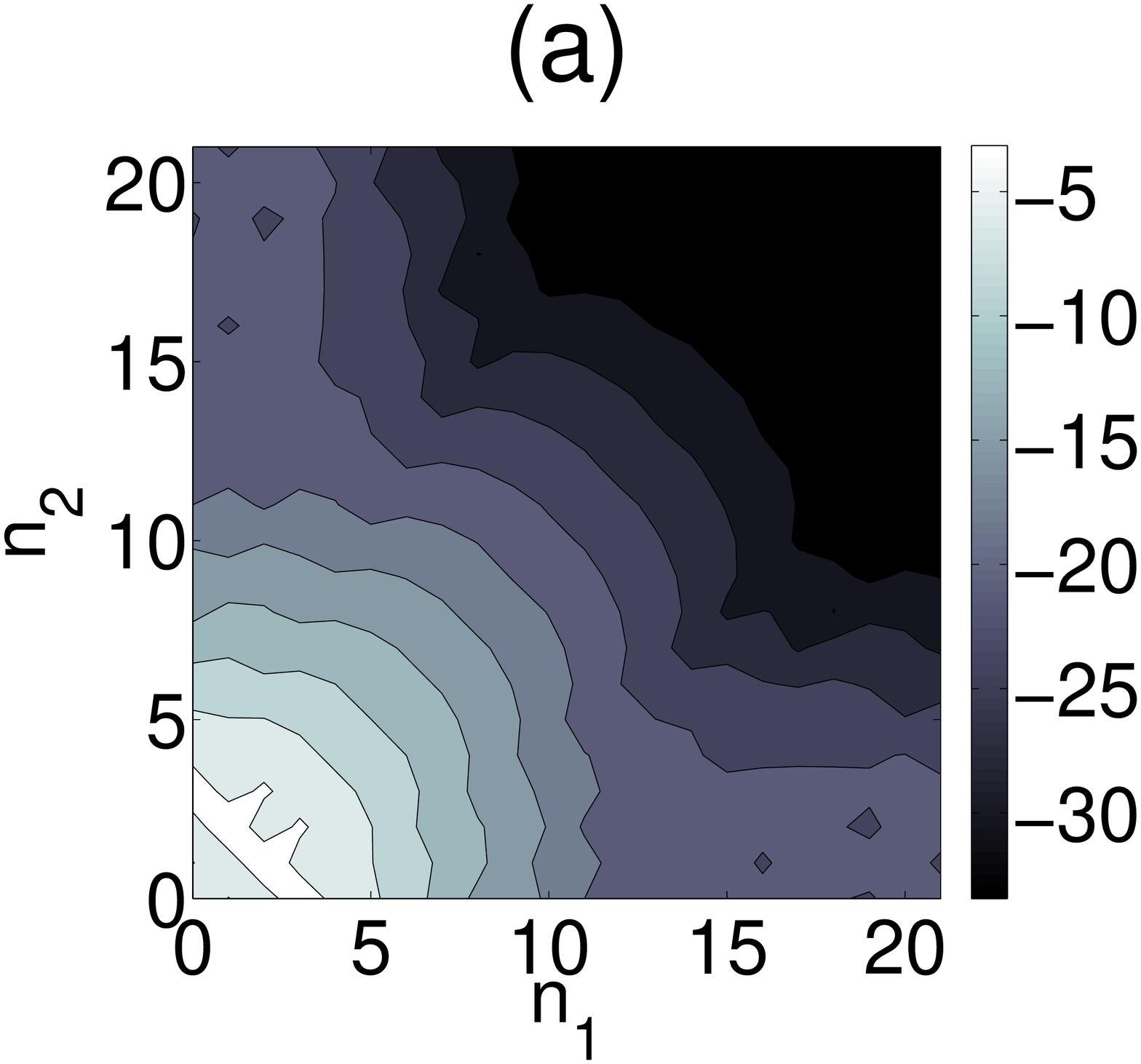}
\includegraphics[width=1.68in]{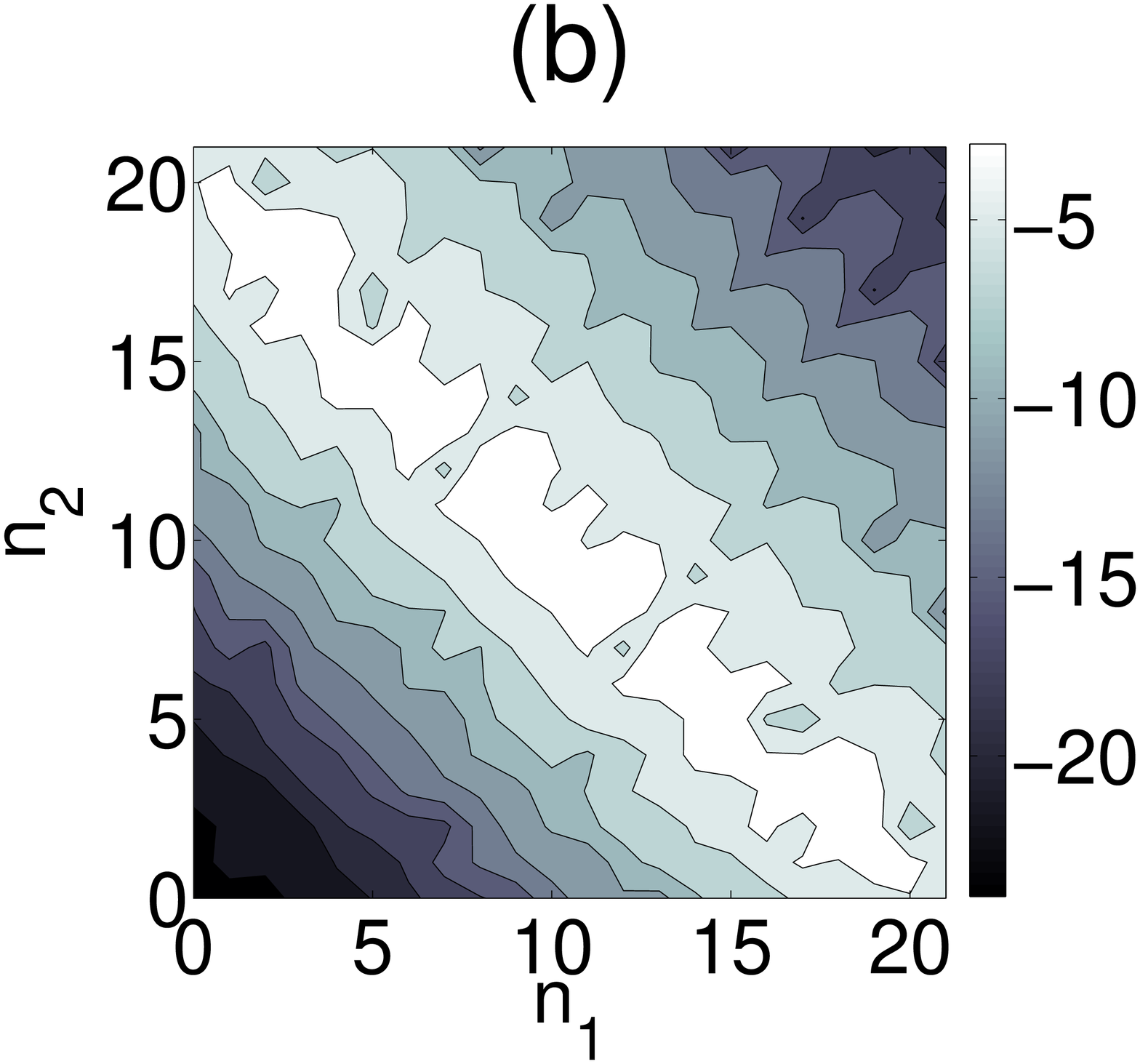}
\includegraphics[width=1.68in]{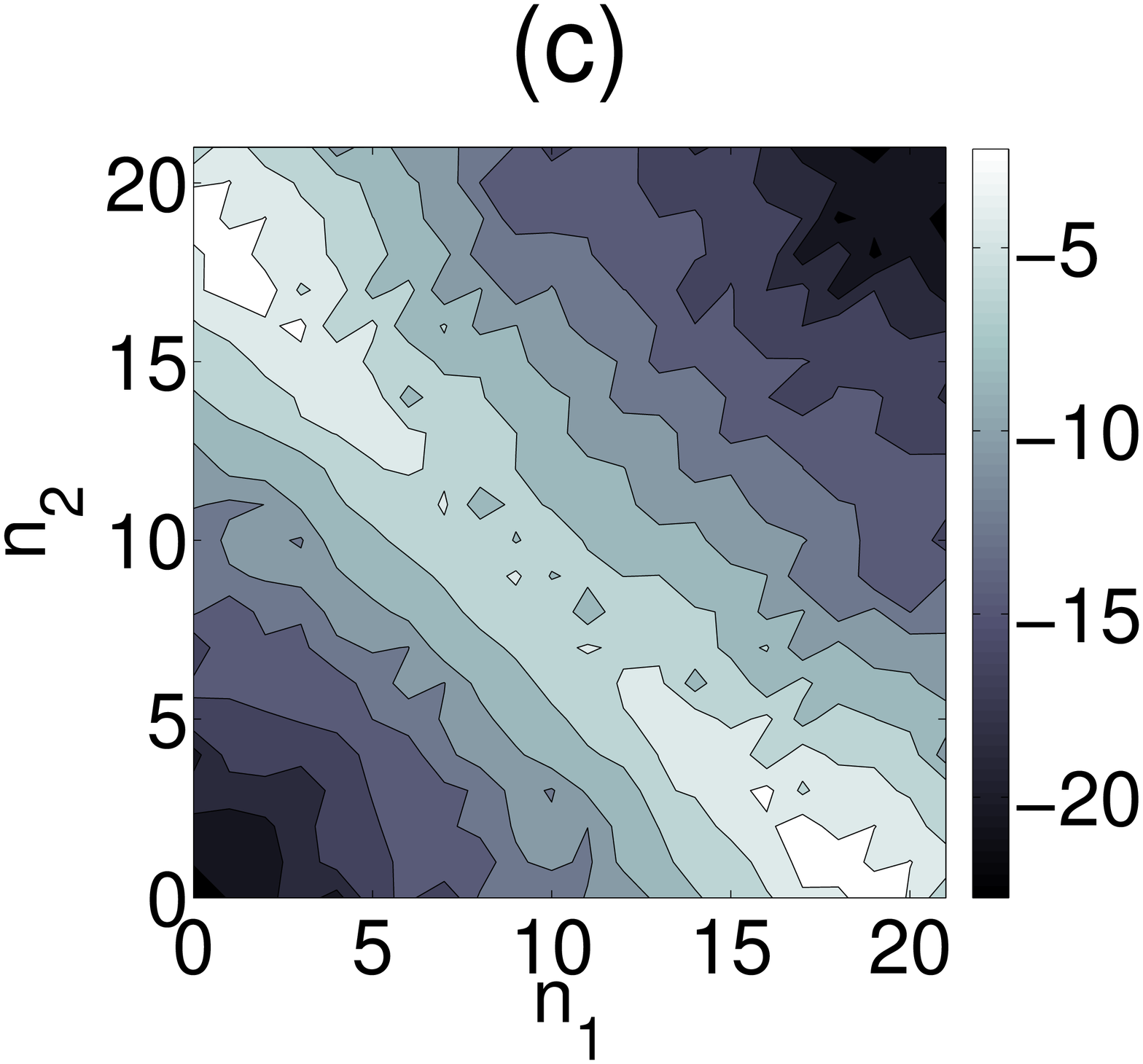}
\includegraphics[width=1.68in]{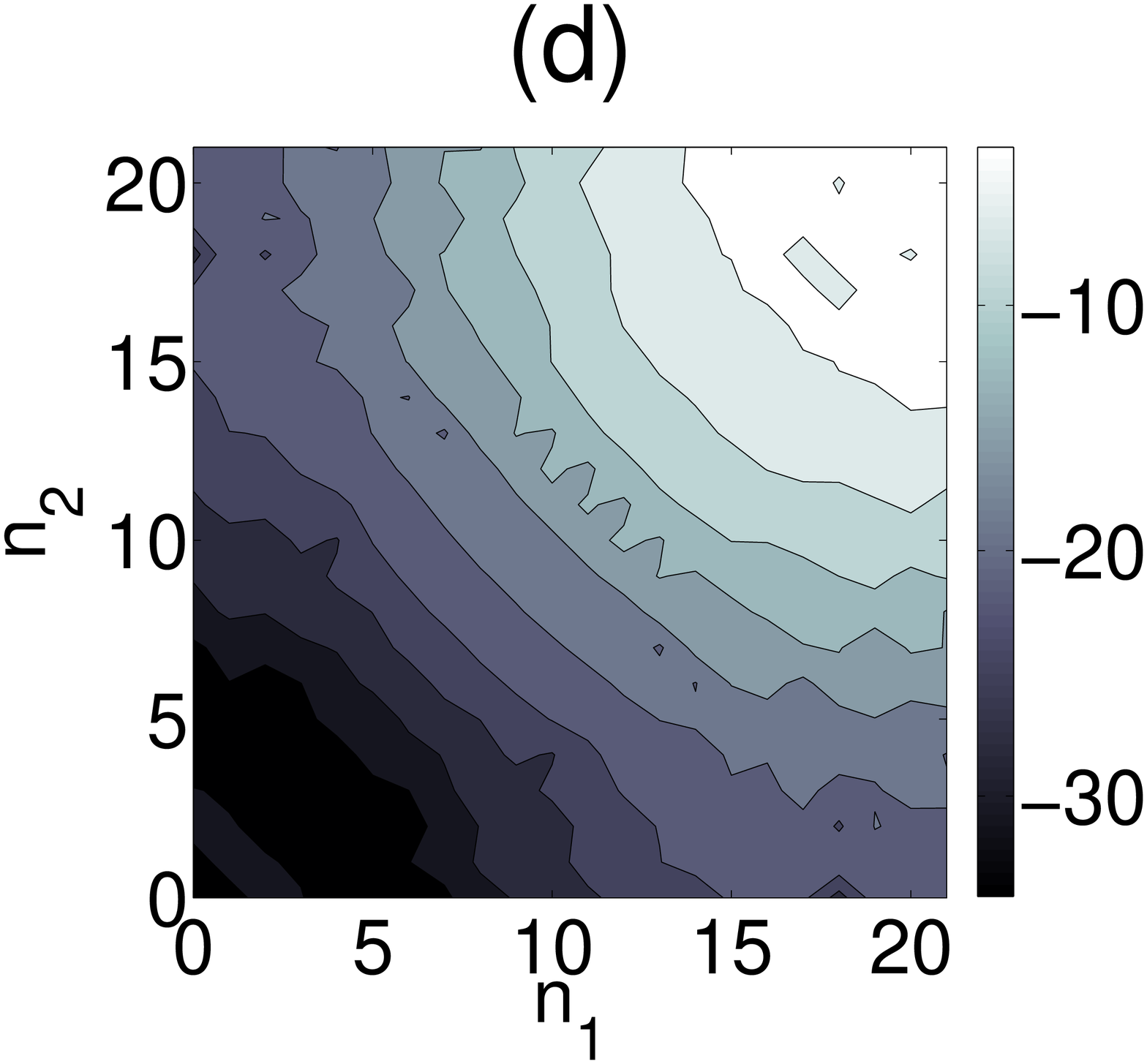}
\caption{\label{entcontour}Contour plots of the logarithm of the density of the
symmetric eigenstates marked by labeled arrows in Fig.\ref{entanglement} for
the case $\zeta = 0.1$:
(a) S-5 (arrow 1), (b) S-117 (arrow 2), (c) S-95 (arrow 3), (d) S-246 (arrow
4). The normalized bias current is $\gamma=0.945$ (22 levels per junction).}
\end{center}
\end{figure} 

In Fig.\ref{entanglement} we show the entanglement of the eigenstates for
different values of the coupling strength $\zeta$. For $\zeta=0$ the
entanglement has the values 0 and 0.5 corresponding to the basis states with
equal and distinct number of quanta at each junction respectively. 
When $\zeta>0$ the eigenstates become linear superpositions of
the basis states and the entanglement rises, being larger as long as more
basis states are involved in building up an eigenstate. This can be seen in
Fig.\ref{entcontour}, where we plot contours of the density of four
symmetric eigenstates marked by labeled arrows in Fig.\ref{entanglement}
for $\zeta=0.1$: the low-energy eigenstate marked by the arrow 1 in
Fig.\ref{entanglement} consists mainly of a
superposition of a few basis states $|n_1,n_2\rangle$ fulfilling
$n_1+n_2=3$ as seen in Fig.\ref{entcontour}-a, hence the entanglement is
relatively small. When going up in energy the entanglement in the eigenstates
quickly increases, becoming maximum in the central part of the energy spectrum
and then decreases. An eigenstate in this region of the spectrum like
the one marked by the arrow 2 in Fig.\ref{entanglement} involves many
basis states fulfilling $n_1+n_2=20$ (Fig.\ref{entcontour}-b), hence the
entanglement is large. However, for QB states living in the same energy region 
like the one marked by the
arrow 3 in Fig.\ref{entanglement}, which has the form shown in
eq. (\ref{eq:QB}) as visible in Fig.\ref{entcontour}-c, the entanglement is smaller and tends to 0.5 as expected. Finally,
high-energy eigenstates like the one marked by the arrow 4 in
Fig.\ref{entanglement} involve not so many basis states
(Fig.\ref{entcontour}-d), therefore the entanglement is also smaller.

We also computed the {\it von Neumann
entropy} \cite{Hines2003} (see appendix for explicit formulas), which is another
standard measure of entanglement, and the results were consistent with those discussed above.

We thus conclude, that when QBs appear in a certain part of the energy spectrum,
their entanglement drops as compared to the typical entanglement of nearby states. The reason is, that QB states predominantly excite two symmetry-related
basis states, as opposed to the typical excitation of many other basis states.

\section{Possible experimental observation of quantum breathers}

The experimental observation of QBs may be possible by using the scheme of
McDermott {\it et al} for simultaneous state measurement of
coupled Josephson phase qubits \cite{Martinis}, where by applying current
pulses in the bias current through each junction the time evolution of the occupation
probabilities in the qubits is measured. By applying a microwave
pulse on one of the junctions we excite it into a high energy
single-junction state with energy $\varepsilon_l$ and leave the other one in the ground state. In this
way we have an initial state similar to the ones shown in Fig.\ref{fig3}-c and d.
After a variable period of time we apply simultaneous current pulses
to the junctions to lower their energy barriers $\Delta U$ and
enhance the probability of tunneling outside the potential well. Then we test which
junction switches to the resistive state (detected by a measurable voltage across it). By
repeating the measuring many
times we obtain the populations in the junctions as a function of the time between the initial
pulse and the simultaneous measuring pulses.

Let us discuss the so far neglected quantum escape. For that	
we computed
$\tau_{escape}$ by using the semiclassical formula \cite{Landau}
\begin{equation}\label{eq:semiclass}
\tau_{escape}^{-1}(\varepsilon) = \frac{\omega(\varepsilon)}{2\pi}\exp\left\{-\frac{2}{\hbar}
\int_a^b p(\varphi) d\varphi\right\},
\end{equation}
where $a$ and $b$ are the turning points of the classical motion in the
reversed potential at $U(\varphi)=\varepsilon$,
$p(\varphi)=\sqrt{2[U(\varphi)-\varepsilon]}$, and $\omega (\varepsilon)/2\pi$ is
the frequency of the oscillations inside the initial well.
In table \ref{times} we show the escape time from different metastable states,
and we compare it with the tunneling time $\tau_{tunnel}$ of an initial excitation
$|\Psi(0)\rangle = |0,l\rangle$ between the two junctions, estimated from the
energy splitting of the (symmetric-antisymmetric) pair of eigenstates with the largest overlap with
the initial excitation. We see that for $l=19$, where we excite QBs, the escape time is long enough
for observing at least one tunneling exchange between the two junctions
before escaping to the resistive state. Note that the cases $l=18$ and 17
also excite QBs which would show even more tunneling exchanges before
escaping. The case $l=16$ does not excite QBs but eigenstates that, though
having small energy splitting, do not show strong site correlation of quanta
as in the previous cases. From these results we expect that escaping to the resistive state will not prevent from the
experimental observation of QB excitations.

\begin{table}[h!]
\caption{Escape times for metastable states in a single Josephson junction
$\tau_{escape}$ estimated by formula (\ref{eq:semiclass}), and the tunneling time
of the initial excitation$|\Psi(0)\rangle = |0,l\rangle$ between the two
junctions $\tau_{tunnel}$ estimated from energy splittings.}
\vspace{0.1cm}
\begin{tabular}{|c|c|c|}
\hline
$l$ & $\tau_{tunnel}$ (ns) & $\tau_{escape}$ (ns) \\ \hline
20 & 348 & 42  \\ \hline
19 & $1.8\times 10^3$ & $3.5\times 10^3$  \\ \hline
18 & $10.16\times 10^3$ & $503.2\times 10^3$  \\ \hline
17 & $2.3\times 10^3$ & $71.2\times 10^6$  \\ \hline
16 & 366 & $1.62\times 10^9$  \\ \hline
\end{tabular}
\label{times}
\end{table}

Another phenomenon that was not taken into account in our quantum model is
decoherence. To be able to observe tunneling between the junctions the coherence time 
has to be longer than the shortest tunneling time between the junctions, which is on the
order of 1 ns in the cases shown in Fig.\ref{fig3}-a and e. In the experiment
shown in \cite{Berkley} using a few levels per junction they obtained a coherence time on
the same order. However, in the experiment in \cite{Martinis} the coherence
time was about 25 ns, and more recently in \cite{Steffen} the coherence time was
approximately 80 ns. We expect that further improvements in experiments
\cite{Steffen2006} will give us even longer coherence times.

Note that the above coherence times are shorter than the tunneling times of QB excitations
(see table \ref{times}), hence decoherence is an effect that can not be
ignored if one wants to do a more realistic quantum description of the
system. When exciting a JJ to high-energy states relaxation (over
dephasing) is usually the main source of decoherence. We can make a crude
estimation of the corresponding relaxation time $T_1$ by
using $T_1 \simeq hQ/\varepsilon_l$ ($Q$ is the quality factor of
the junctions), which holds for a harmonic
potential \cite{Martinis1987,Esteve1986}. For $l=19,18$ and 17,
$\varepsilon_l/h$ is around 150 GHz (see Fig.\ref{fig2}-b). For the JJs used in
\cite{Steffen2006}, $Q$ is between 500 and 1000, which leads to a relaxation time
between 3 ns and 6 ns. It is much smaller than the tunneling time of the
QB excitations, therefore one would expect to see instead of
tunneling, a freezing of the QBs on one of the junctions before they decohere due to relaxation. 

One could obtain more feasible results by increasing the bias current in such a way
that there are less energy levels in the junctions. With this, exciting a QB would
need less energy, and the relaxation time becomes longer. The
tunneling time of that QB excitation is shorter, and might be even shorter than the relaxation time, allowing one to observe
tunneling before relaxation. This possibility, and the inclusion of
decoherence in our model, are issues that will be addressed in a future work.

\section{Conclusions}

We have studied the classical and quantum dynamics of high-energy localized
excitations in a system of two capacitively coupled JJs. In
the classical case the equations of motion admit time periodic localized
excitations (discrete breathers) which can be numerically computed. For the
quantum case we showed that excitation of one of the junctions to a high level,
leaving the another junction in the ground state, may strongly overlap with QBs
(tunneling-pairs eigenstates) that live in the central part of the
energy spectrum and localize energy on one of the junctions for a long
(tunneling) time.
This result would not
qualitatively change if we excite a (coherent or incoherent) superposition of several product basis
states instead of only one. By using the density function for asymmetric
superpositions of QB states one can
realize how many quanta can be excited at each junction in order to excite QB states.

In addition to what was described above, the system showed other
interesting properties: We found that the system nearly conserves the
total number of quanta, which comes from the fact that the coupling between
the junctions couples just slightly eigenstates components with different
total number of quanta. This opens the possibility to study larger systems
without too big computational cost. When computing the fluctuation in the total number of
quanta for each eigenstate, QB states show the smallest
fluctuations. We showed that entanglement, which
reflects how many basis states have significant weight in an eigenstate,
increases with energy in most of the eigenstates, becoming maximum at the center of the
spectrum and then decreases. 
QB states from the same energy region are less entangled. 
This is because a QB state mainly consists of a symmetric
or antisymmetric combination of two product basis states localizing many quanta on one of the
junctions.

With the available
techniques for manipulating Josephson-junction qubits the experimental observation of
QB excitations is possible. Escaping to the resistive state of the junctions
(which together with decoherence was not taken into account in our quantum model) would not
prevent us from doing that, and we expect that improvements on preparation
(higher quality factors) and isolation techniques of JJs make it 
possible to reach long enough coherence times
in order to observe the phenomena we described in this work. By changing the
parameters of the system (bias current and coupling strength) one could vary
the energy, and hence the tunneling time of a QB excitation with respect to
the coherence time, in such a way that it becomes larger than that tunneling
time.


\acknowledgments

We thank A. Ustinov, J. Lisenfeld, and T. Ohki for useful discussions. This work was supported
by the DFG (grant No. FL200/8) and by the ESF network-programme AQDJJ.

\appendix*

\section{Measures of entanglement}

Let $\mathcal{X}\neq 0$ be a real eigenstate written in a basis
of product states $\{|n_1,n_2\rangle \}= \{|n_1\rangle\otimes |n_2\rangle\}$
with $n_1,n_2=1,\ldots N$. It is a $N\times N$ matrix with elements
$\chi_{n_1,n_2}=\langle n_1,n_2|\chi\rangle$.
We define the geometric measure of
entanglement of this eigenstate by the following quantity:
\begin{equation}\label{eq:delta}
\Delta = \sum_{n_1,n_2}^N(\chi_{n_1,n_2} - f_{n_1}g_{n_2})^2,
\end{equation}
Where the vector functions $\mathbf{f}=(f_1,\ldots,f_{n_1},\ldots)^t$ and $\mathbf{g}=(g_1,
\ldots,g_{n_2},\ldots)^t$ are such that $\Delta$ is minimum.
The quantity $\Delta$ measures how far the eigenstate is from being a product of functions
depending respectively on the numbers $n_1$ and $n_2$.

The
minimization of $\Delta$ leads to the formula \cite{geoent1,geoent2,geoent3}:
\begin{equation}\label{eq:ent}
\Delta = \| \mathcal{X}^{\mu} \|^2 - \lambda_{max},
\end{equation}
where $\mathcal{X}^{\mu}$ is a $N\times N$ matrix with elements
$\chi_{n_1,n_2}^{\mu}$, $\lambda_{max}$ is the maximum eigenvalue of the
$N\times N$ reduced density matrix 
\begin{equation}\label{eq:Amatrix}
\mathcal{A} =
\mathcal{X}^{\mu}(\mathcal{X}^{\mu})^t,
\end{equation}
 and $\| \mathcal{X}^{\mu} \|^2 =
\sum_{n_1,n_2}^N(\chi_{n_1,n_2}^{\mu})^2$.

Another standard measure of entanglement in the eigenstates is
the {\it von Neumann entropy}, which is used in
information theory \cite{Hines2003}:
\begin{eqnarray}
S(\hat{\rho}_1) &=& -Tr\{\hat{\rho}_1\log_2\hat{\rho}_1\}, \\
\label{eq:entro2} &=& -\sum_k \lambda_k\log_2(\lambda_k),
\end{eqnarray}
where $\log_2$ refers to the logarithm taken in base 2. $\hat{\rho}_1$ is the 
reduced density operator of either of the subsystems:
\begin{equation}
\hat{\rho}_1 = Tr_2(\hat{\rho}),
\end{equation}
where $Tr_2$ is the partial trace over the subsystem 2.
$\{\lambda_k\}$ is the set of eigenvalues of the reduced density
operator $\hat{\rho}_1$.

For the system of coupled JJs an eigenstate has the form:
\begin{equation}
|\chi\rangle = \sum_{n_1,n_2}^N \chi_{n_1,n_2} |n_1,n_2\rangle.
\end{equation}
Then the density operator is
\begin{eqnarray}
\hat{\rho} &=& |\chi\rangle\langle\chi| \\
&=& \sum_{n_1,n_2}^N \sum_{n_1^{\prime},n_2^{\prime}}^N
\chi_{n_1,n_2}\chi_{n_1^{\prime},n_2^{\prime}}^* |n_1,n_2\rangle\langle
n_1^{\prime},n_2^{\prime}|,
\end{eqnarray}
hence the reduced density operator is
\begin{eqnarray}
\hat{\rho}_1 &=& Tr_2(\hat{\rho}) \\
&=& \sum_{n_1,n_1^{\prime}}^N \left\{ \sum_{n_2}^N \chi_{n_1,n_2}\chi_{n_1^{\prime},n_2}^*\right\}
|n_1\rangle\langle n_1^{\prime}|.
\end{eqnarray}
The matrix elements of this reduced operator are
\begin{eqnarray}
\langle n|\hat{\rho}_1|m\rangle &=& \sum_{n_2}^N \chi_{n,n_2}\chi_{m,n_2}^*, \\
& = & A_{n,m},
\end{eqnarray}
where $A_{n,m}$ is a matrix element of $\mathcal{A}$ defined in
eq. (\ref{eq:Amatrix}). To compute the von Neumann entropy one diagonalizes this matrix and uses eq. (\ref{eq:entro2}).

Despite the fact that we found similarity in the variation of the two measures
when studying QB eigenstates, it is interesting to note, that monotonicity
does not hold in general, i.e. if one measure is telling that a given state is stronger
entangled than another one, that property may be reversed when using the 
other measure. The geometric measure
is an unambiguous number of the shortest distance from a given state to the
subspace of product states, it allows to reconstruct the optimum product
state, and it has a clear relation to the Hartree approximation \cite{geoent1}.
For these reasons we presented the numerical results using the geometric measure,
rather than the entropy measure.



\begin{thebibliography}{99}

\bibitem{Likharev}
Likharev, K. K. {\it Dynamics of Josephson junctions and circuits} (Gordon and
Breach Science Publishers, Philadelphia, 1984).

\bibitem{Leggett}{\it Quantum Computing and Quantum Bits in Mesoscopic
Systems}, edited by A. Leggett, B. Ruggiero, and P. Silvestrini (Kluwer
Academic/Plenum Publishers, New York, 2004).

\bibitem{Esteve}{\it Les Houches 2003, Quantum Entanglement and Information
Processing}, edited by D. Est\`eve, J. M. Raimond, and J. Dalibard (Elsevier,
Amsterdam, 2004).

\bibitem{Martinis2}J. M. Martinis, M. H. Devoret, and J. Clarke,
Phys. Rev. Lett. {\bf 55}, 1543 (1985).

\bibitem{Steffen2006}M. Steffen {\it et al} Phys. Rev. Lett. {\bf 97}, 050502
(2006).

\bibitem{Martinis}R. McDermott {\it et al}, Science {\bf 307}, 1299 (2005).

\bibitem{Steffen}M. Steffen {\it et al}, Science {\bf 313}, 1423 (2006). 

\bibitem{Graham}
R. Graham, M. Schlautmann, and D. L. Shepelyansky, Phys. Rev. Lett.
{\bf 67}, 255 (1991).

\bibitem{Montangero}
S. Montangero, A. Romito, G. Benenti, and R. Fazio, Europhys. Lett. {\bf 71},
893 (2005).

\bibitem{Pozzo}
E. N. Pozzo, and D. Dom\'inguez, Phys. Rev. Lett. {\bf 98}, 057006 (2007).

\bibitem{physicstoday}
D. K. Campbell, S. Flach, Y. S. Kivshar, {\it Physics Today} p. 43, January
2004.

\bibitem{FlachPhysRep295}
S. Flach, C.R. Willis, Phys. Rep. {\bf 295}, 181 (1998).

\bibitem{Sievers}
A. J. Sievers, J. B. Page, in: G. K. Horton, A. A. Maradudin (eds.), {\it
Dynamical Properties of Solids VII, Phonon Physics. The Cutting Edge},
Elsevier, Amsterdam (1995), p. 137.

\bibitem{AubryPhysicaD103}
S. Aubry, Physica D {\bf 103}, 201 (1997).

\bibitem{SchwarzPRL83}
U. T. Schwarz, L. Q. English, A. J. Sievers, Phys. Rev. Lett. {\bf 83}, 223
(1999).

\bibitem{SatoNature}
M. Sato, A. J. Sievers, Nature {\bf 432}, 486 (2004).

\bibitem{SwansonPRL82}
B. I. Swanson, J. A. Brozik, S. P. Love, G. F. Strouse, A. P. Shreve,
A. R. Bishop, W.-Z. Wang, M. I. Salkola, Phys. Rev. Lett. {\bf 82}, 3288
(1999).

\bibitem{TriasPRL84}
E. Trias, J. J. Mazo, T. P. Orlando, Phys. Rev. Lett. {\bf 84}, 741 (2000).

\bibitem{BinderPRL84}
P. Binder, D. Abraimov, A. V. Ustinov, S. Flach, Y. Zolotaryuk,
Phys. Rev. Lett. {\bf 84}, 745 (2000).

\bibitem{EisenbergPRL81}
H. S. Eisenberg, Y. Silberberg, R. Morandotti, A. R. Boyd, J. S. Aitchison,
Phys. Rev. Lett. {\bf 81}, 3383 (1998).

\bibitem{FleischerNature422}
J. W. Fleischer, M. Segev, N. K. Efremidis, D. N. Christodoulides, Nature {\bf
422}, 147 (2003).

\bibitem{SatoPRL90}
M. Sato, B. E. Hubbard, A. J. Sievers, B. Ilic, D. A. Czaplewski,
H. G. Craighead, Phys. Rev. Lett. {\bf 90}, 044102 (2003).

\bibitem{EiermannPRL92}
B. Eiermann, Th. Anker, M. Albiez, M. Taglieber, P. Treutlein, K.-P. Marzlin,
M. K. Oberthaler, Phys. Rev. Lett. {\bf 92}, 230401 (2004).

\bibitem{Fleurov}V. Fleurov, Chaos {\bf 13}, 676 (2003).

\bibitem{ScottPhysLettA119}
A. C. Scott, J. C. Eilbeck, Phys. Lett. A {\bf 119}, 60 (1986).

\bibitem{BernsteinNonlin3}
L. Bernstein, J. C. Eilbeck, and A. C. Scott, Nonlinearity {\bf 3}, 293
(1990).

\bibitem{BernsteinPhysicaD68}
L. J. Bernstein, Physica D {\bf 68}, 174 (1993).

\bibitem{WrightPhysicaD69}
E. Wright, J. C. Eilbeck, M. H. Hays, P. D. Miller, and A. C. Scott, Physica D
{\bf 69}, 18 (1993).

\bibitem{Eilbeck94}
A. C. Scott, J. C. Eilbeck, and H. Gilh\o j, Physica D {\bf 78}, 194 (1994).

\bibitem{Wang}W. Z. Wang, J. T. Gammel, A. R. Bishop, and M. I. Salkola,
Phys. Rev. Lett. {\bf 76}, 3598 (1996).

\bibitem{Aubry}
S. Aubry, S. Flach, K. Kladko, and E. Olbrich, Phys. Rev. Lett. {\bf 76}, 1607
(1996).

\bibitem{Flach1}
S. Flach and V. Fleurov, J. Phys.: Cond. Matt. {\bf 9}, 7039 (1997).

\bibitem{Fleurov1998}
V. Fleurov, R. Schilling, and S. Flach, Phys. Rev. E {\bf 58}, 339 (1998).

\bibitem{kalosakas2}
G. Kalosakas, A. R. Bishop, V. M. Kenkre, J. Phys. B {\bf
36}, 3233 (2003).

\bibitem{Dorignac2004}
J. Dorignac, J. C. Eilbeck, M. Salerno, and A. C. Scott, Phys. Rev. Lett. {\bf
93}, 025504 (2004).

\bibitem{Eilbeck2004}
J. C. Eilbeck and F. Palmero, Phys. Lett. A {\bf 331}, 201 (2004).

\bibitem{Pinto}R. A. Pinto and S. Flach, Phys. Rev. A {\bf 73},022717 (2006).

\bibitem{Schulman}L. S. Schulman, D. Tolkunov, and E. Mihokova,
Phys. Rev. Lett. {\bf 96}, 065501 (2006).

\bibitem{Schulman2006}
L. S. Schulman, D. Tolkunov, and E. Mih\'okov\'a, Chem. Phys. {\bf 322}, 55 (2006).

\bibitem{Proville2006}
L. Proville, Physica D {\bf 216}, 191 (2006).

\bibitem{Ivic2006}
Z. Ivi\'c, G. P. Tsironis, Physica D {\bf 216}, 200 (2006).

\bibitem{Eilbeck03}
J. C. Eilbeck, in:
Localization and Energy Transfer in Nonlinear Systems, Ed. L. Vazquez,
R. S. MacKay and M. P. Zorzano, p.177 (World Scientific, Singapore 2003).

\bibitem{Fillaux1990}F. Fillaux, C. J. Carlile, Phys. Rev. B {\bf 42}, 5990
(1990).

\bibitem{Fillaux1998}F. Fillaux, C. J. Carlile, G. J. Kearley, Phys. Rev. B
{\bf 58}, 11416 (1998).
 
\bibitem{Richter1988}
L. J. Richter, T. A. Germer, J. P. Sethna, and W. Ho, Phys. Rev. B {\bf 38},
10403 (1988).

\bibitem{GuyotSionnest1991}
P. Guyot-Sionnest, Phys. Rev. Lett. {\bf 67}, 2323 (1991).

\bibitem{Dai1994}
D. J. Dai, and G. E. Ewing, Surf. Sci. {\bf 312}, 239 (1994).

\bibitem{Chin1995}
R. P. Chin, X. Blase, Y. R. Shen, and S. G. Louie, Europhys. Lett. {\bf 30},
399 (1995).

\bibitem{Jakob1996}
P. Jakob, Phys. Rev. Lett. {\bf 77}, 4229 (1996).

\bibitem{JakobPr75}
P. Jakob, Appl. Phys. A: Mater. Sci. Process. {\bf 75}, 45 (2002).

\bibitem{Okuyama2001}
H. Okuyama, T. Ueda, T. Aruga, and M. Nishijima, Phys. Rev. B {\bf 63}, 233404
(2001).

\bibitem{Edler2004}
J. Edler, R. Pfister, V. Pouthier, C. Falvo, and P. Hamm,
Phys. Rev. Lett. {\bf 93}, 106405 (2004).

\bibitem{Pinto2007}R. A. Pinto and S. Flach, Europhys. Lett. {\bf 79}, 66002 (2007).

\bibitem{Berkley}A. J. Berkley, H. Xu, R. C. Ramos, M. A. Gubrud,
F. W. Strauch, P. R. Johnson, J. R. Anderson, A. J. Dragt, C. J. Lobb, and
F. C. Weestood, Science {\bf 300}, 1548 (2003).

\bibitem{Johnson}P. R. Johnson, F. W. Strauch, A. J. Dragt, R. C. Ramos,
C. J. Lobb, J. R. Anderson, and F. C. Wellstood, Phys. Rev. B {\bf 67}, 020509(R)
(2003).

\bibitem{Blais}A. Blais, A. Maassen van den Brink, and A. M. Zagoskin,
Phys. Rev. Lett. {\bf 90}, 127901 (2003).

\bibitem{Flach2}{\it Computational studies of discrete breathers}, S. Flach,
in {\it Energy Localization and Transfer}, Edited by T. Dauxois, Anna
Litvak-Hinenzon, Robert MacKay (University of Warwick, UK), and Anna Spanoudaki, (World
Scientific, 2004), p1.

\bibitem{Aubry1}D. Chen, S. Aubry, and G. P. Tsironis, Phys. Rev. Lett. {\bf
77}, 4776 (1996).

\bibitem{Fourier}C. Clay Marston and G. G. Balint-Kurti, J. Chem. Phys. {\bf
91}, 3571 (1989).

\bibitem{harmonic}By harmonic approximation we mean that we use the equations
(\ref{eq:number}) and (\ref{eq:momenta}) for the number and momentum
operators, which hold for the harmonic oscillator potential. For the
anharmonic potential of the Josephson junctions these relations do not necessarily
hold.

\bibitem{Davis1981JChemPhys75}
M. Davis and E. Heller, J. Chem. Phys. {\bf 75}, 246 (1981).

\bibitem{Keshavamurthy2007IntRevPhysChem26}
S. Keshavamurthy, Int. Rev. Phys. Chem. {\bf 26}, 521 (2007).

\bibitem{Keshavamurthy2005JChemPhys122}
S. Keshavamurthy, J. Chem. Phys. {\bf 122}, 114109 (2005).


\bibitem{geoent1}T.-C. Wei and P. M. Goldbart, Phys. Rev. A {\bf 68},
042307 (2003).

\bibitem{geoent2}A. Shimony, Ann. NY. Acad. Sci. {\bf 755}, 675 (1995).

\bibitem{geoent3}H. Barnum and N. Linden, J. Phys. A.: Math. Gen. {\bf 34},
6787 (2001).


\bibitem{Tonel2005}A. P. Tonel, J. Links, and A. Foerster, J. Phys. A:
Math. Gen. {\bf 38}, 1235 (2005).

\bibitem{Fu2006}L. B. Fu and J. Liu, Phys. Rev. A {\bf 74}, 063614 (2006).

\bibitem{Hines2003}
A. P. Hines, R. H. McKenzie, and G. J. Milburn, Phys. Rev. A {\bf 67}, 013609 (2003).
\bibitem{Landau}
Landau, L. D. and Lifshitz, E. M. {\it Quantum Mechanics} (Pergamon London, 1958).

\bibitem{Martinis1987}
J. M. Martinis, M. H. Devoret, and J. Clarke, Phys. Rev. B {\bf 35}, 4682 (1987).

\bibitem{Esteve1986}
D. Esteve, M. H. Devoret, and J. M. Martinis, Phys. Rev. B {\bf 34}, 158 (1986).



\end{thebibliography}
\end{document}